\documentclass[letterpaper,twocolumn,10pt]{article}
\usepackage{usenix2019_v3}

\newcommand{\AttackName}{Fallout\xspace}
\title{\Large \bf \AttackName: Reading Kernel Writes From User Space}

\usepackage{authblk}
\author[1]{Marina Minkin}
\author[2]{Daniel Moghimi}
\author[3]{Moritz Lipp}
\author[3]{Michael Schwarz}
\author[4]{Jo Van Bulck}
\author[1]{Daniel Genkin}
\author[3]{Daniel Gruss}
\author[4]{Frank Piessens}
\author[2]{Berk Sunar}
\author[5]{Yuval Yarom}
\affil[1]{University of Michigan}
\affil[2]{Worcester Polytechnic Institute}
\affil[3]{Graz University of Technology}
\affil[4]{imec-DistriNet, KU Leuven}
\affil[5]{The University of Adelaide and Data61}

\usepackage{enumitem}
\usepackage{tikz}
\usetikzlibrary{arrows}
\usepackage{lipsum}
\usepackage{authblk}
\usepackage{hyperref}
\usepackage{xcolor}
\usepackage{listings}
\usepackage{xspace}
\usepackage[nameinlink,capitalise]{cleveref}
\usepackage{etoolbox}
\usepackage{pifont}
\usepackage{colortbl}
\usepackage{graphicx}

\newcommand{\cmark}{\textcolor{mGreen}{\ding{51}}\xspace}%
\newcommand{\xmark}{\textcolor{red}{\ding{55}}\xspace}%

\newcommand\ccmark{%
	\tikz\node[circle,draw=mGreen,line width=0.25mm,inner sep=0.1pt]{\cmark};\xspace%
}
\newcommand\cxmark{%
	\tikz\node[circle,draw=red,line width=0.25mm,inner sep=0.5pt]{\xmark};\xspace%
}
\newcommand{\textccmark}{\textcolor{mGreen}{\textcircled{\cmark}}\xspace}
\newcommand{\textcxmark}{\textcolor{red}{\textcircled{\xmark}}\xspace}
\newcommand{\ie}{i.e.,\xspace} 
\newcommand{\eg}{e.g.,\xspace} 
\newcommand{\FlushReload}{Flush+\allowbreak Reload\xspace}

\newcommand{\PrimeProbe}{Prime+\allowbreak Probe\xspace}
\newcommand{\muop}{$\mu$OP\xspace}
\newcommand{\muops}{$\mu$OPs\xspace}
\newcommand{\instr}[1]{\textsc{#1}\xspace}

\newcommand{\parhead}[1]{\vspace{1pt plus 2pt minus 1pt}\par\noindent\textbf{#1}\hspace{1em plus 0.5em minus 0.5em}}

\hypersetup{
	colorlinks,
	linkcolor={red!50!black},
	citecolor={blue!50!black},
	urlcolor={blue!80!black}
}

\definecolor{mGreen}{rgb}{0,0.6,0}
\definecolor{mGray}{rgb}{0.5,0.5,0.5}
\definecolor{mPurple}{rgb}{0.58,0,0.82}
\definecolor{backgroundColour}{rgb}{0.95,0.95,0.92}
\definecolor{tabledark}{rgb}{0.9, 0.9, 0.9}
\definecolor{tablehighlight}{HTML}{FFB5B5}
\definecolor{darkgreen}{rgb}{0,0.65,0}

\newcommand{\colorcomment}[2]{\leavevmode\unskip\space{\color{#1}[#2]}\xspace}

\newcommand{\refs}{\colorcomment{red}{REFS}}

\lstdefinestyle{customc}
{
	breaklines=true,
	xleftmargin=2em,
	captionpos=b,                    
	language=C,
	showstringspaces=false,
	basicstyle=\footnotesize\ttfamily,
	keywordstyle=\bfseries\color{black},
	commentstyle=\itshape\color{gray!70!black},
	identifierstyle=\color{black},
	stringstyle=\color{red!70!black},
	emph={static,volatile,double,float,signed,unsigned,int,void,size_t,char,for,PAGE_SIZE,if},
	emphstyle={\bfseries\color{blue}},
	numbersep=5pt,
	numbers=left,
	otherkeywords={PAGE_SIZE},
	morekeywords={PAGE_SIZE},
	frame=single,
}

\begin{document}
\maketitle

\pagestyle{empty}

\begin{abstract}
  Recently, out-of-order execution, an important performance optimization
  in modern high-end processors, has been revealed to pose a significant security
  threat, allowing information leaks across security domains.
  In particular, the Meltdown attack leaks information from the operating system
  kernel to user space, completely eroding the security of the system.
  To address this and similar attacks, without incurring the performance costs of
  software countermeasures, Intel includes hardware-based defenses in its
  recent Coffee Lake R processors.

  In this work, we show that the recent hardware defenses are not sufficient.
  Specifically, we present \emph{\AttackName}, a new transient execution attack
  that leaks information from a previously unexplored microarchitectural component called 
  the \emph{store buffer}.  We show how unprivileged user processes can exploit \AttackName
  to reconstruct privileged information recently written by the kernel.  We further show how \AttackName
  can be used to bypass kernel address space randomization.
  Finally, we identify  and explore microcode assists as a hitherto ignored
  cause of transient execution.

  \AttackName affects all processor generations we have tested. 
  However, we notice a worrying regression, where the newer 
  Coffee Lake R processors are more vulnerable to
  \AttackName than older generations.
\end{abstract}

\section{Introduction}

The architecture and security communities will remember 2018 as the year of
Spectre~\cite{Kocher2019spectre} and Meltdown~\cite{Lipp2018meltdown}.
Speculative and out-of-order execution, 
which have been considered for decades to be harmless
and valuable performance features, were discovered to have dangerous
industry-wide security implications, affecting operating systems
(OSs)~\cite{Lipp2018meltdown, Kocher2019spectre},
browsers~\cite{Kocher2019spectre, specrev4}, virtual
machines~\cite{weisse2018foreshadowNG}, trusted execution environments (e.g.,
SGX)~\cite{Vanbulck2018foreshadow}, AES hardware
accelerators~\cite{stecklina2018lazyfp} and more.

Meltdown, in particular, is a severe 
hardware issue. In a Meltdown attack, an unprivileged attacker performs an
explicit access violation to a privileged memory location containing the OS's
kernel. The CPU responds with the value from that address, while
marking the load operation as faulty. Perhaps most shockingly, the CPU then
allows subsequent transient computation on the returned value.
Finally, by the time that the CPU recognizes the violation
and attempts to undo the damage caused by transient execution, the attacker
already had sufficient cycles to leak the kernel data using a microarchitectural 
covert channel, such as via the processor's cache~\cite{GeYCH18,Maurice2017Hello}.

Recognizing the danger posed by this hardware issue, the computer industry
mobilized.  Potentially incurring significant performance losses~\cite{Gregg2018kpti}, all major OS deployed countermeasures based on the KAISER patch~\cite{Gruss2017KASLR}, which removes the mapping of
kernel pages from the address space of user processes.  At a high level, Kernel Page Table Isolation (KPTI)
relies on the idea that even if the attacker can access the entire currently
mapped address space, the attacker lacks the capabilities of accessing memory outside of the current address space, thus leaving the kernel safely out of reach. 

Unfortunately, with Foreshadow~\cite{Vanbulck2018foreshadow} and
Foreshadow-NG~\cite{weisse2018foreshadowNG} it became clear that an attacker can
transiently access even pages that are not mapped into the address space. The attacker then
subsequently exploits a Meltdown-like technique  
to leak privileged data, including enclave secrets safeguarded by
Intel's Software Guard eXtensions (SGX)~\cite{Vanbulck2018foreshadow} or across
virtual machines running on the same physical
host~\cite{weisse2018foreshadowNG}. 

In an attempt to claw back some of the performance loss, and to
permanently eliminate Foreshadow and Meltdown related issues,
Intel announced already back in 2018 strong, silicon-based Meltdown defenses in future processors
enumerating Rogue Data Cache Load resilience (RDCL\_NO)~\cite{IntelMitigations}.
With the recent release of the 9th generation Coffee Lake R microarchitecture,
such Meltdown-resistant processors are finally available on the mass consumer market.
The RDCL\_NO security feature promises to obviate the need for KPTI
and other defenses, while improving overall performance~\cite{i9eval}.
However, while Intel claims that these fixes address Meltdown and Foreshadow,
it remains unclear  whether new generations of Intel processors are properly
protected against Meltdown-type transient execution attacks.  
Thus, in this work we set out
to investigate the following question:

{\medskip
\centering
 \emph{Is kernel data safe in the new generation of processors?
 Can ad-hoc software
 mitigations be safely disabled on post-Meltdown Intel hardware? }
}

\medskip\noindent

\subsection{Our Contribution}
Unfortunately, in this paper, we answer these questions in the negative.
We present \emph{\AttackName}, a new attack on the hardware-based memory
isolation mechanisms in Intel CPUs.  Using \AttackName, user-space programs can read data that has recently been written by the kernel, as well as 
derandomize Kernel Address Space Layout Randomization (KASLR).  Similarly to
previous transient execution attacks,  \AttackName does not require any
privileges except for the ability to run code, and does not exploit any kernel
vulnerabilities.

\parhead{The Mechanism Behind \AttackName.}
\AttackName exploits an optimization that we call 
\emph{Write Transient Forwarding} (WTF), which incorrectly passes values from
memory writes to subsequent memory reads.
In a nutshell, when the program writes a value to memory,
the processor needs to first translate the virtual address of the
destination to a physical address and then acquire exclusive access
to the location.
Rather than stalling the store instruction and subsequent computation,
the processor records the value and the address in the \emph{store buffer},
and continues executing the program.
The store buffer then resolves the address, acquires the access to the memory
location and stores the data.

When a value is in the store buffer, care should be taken that subsequent loads
from the same address do not read stale values from memory.
To solve this, the processor matches the addresses of all load instructions
against addresses in the store buffer.
In the case of a match, the processor \emph{forwards} the matching value
from the store buffer to the load instruction.
To increase efficiency, the processor uses partial address matches
to rule out the need for store-to-load forwarding.
WTF kicks in when a load instruction partially matches
a preceding store and the processor determines that the load is bound to fail.
In such cases, instead of cleaning up the state of the processor,
it marks the load as faulty, and \emph{incorrectly} forwards the
value of the partially matched store.

\parhead{Exploiting the WTF optimization.}
\AttackName exploits this behavior to leak, through a microarchitectural channel,
the value that WTF incorrectly forwards. The attacker deliberately
performs a faulty load, causing the CPU to transiently forward an incorrect
value from the store buffer. We subsequently leak the value using a
Flush+Reload~\cite{yarom2014flush+} side channel. As the store buffer is a
shared resource used by all software running on a CPU core, the incorrectly-forwarded
value might not even belong to the attacker's process.
Empirically demonstrating this, in this paper, we show how to exploit the WTF
optimization to leak values recently written by the kernel from user space as
well as how to derandomize the kernel's ASLR.

\parhead{\AttackName vs.\ Meltdown}
Like all Meltdown-type attacks, \AttackName exploits transient 
execution past an exception.
However, unlike previous Meltdown-type attacks,
in \AttackName the adversary does not read from the address of the protected
value.  
Instead, the value leaks while the adversary loads from an unrelated memory
address.
As a result, the hardware countermeasures for Meltdown and Foreshadow in
recent Intel processors do not protect against \AttackName.
Finally, we note a worrying regression in recent Intel processors,
where, possibly due to the added hardware countermeasures, 
newer processors seem more vulnerable to \AttackName than previous
generations.

\parhead{Security Analysis of Speculation Mechanisms and Coffee Lake Refresh.}
We present the first analysis of various
exception-creation and exception-suppression mechanisms used to mount
\AttackName across various Intel architectures. As we show, not all creation
and suppression mechanisms are interchangeable, and the exact combination is,
in fact, architecture dependent. Finally, we show that the hardware change in
exception creation and suppression introduced by Intel in the latest Coffee
Lake Refresh architecture make them more vulnerable to our attack.

\parhead{Exploiting Microcode Assists.}
As a final contribution, we identify a hitherto unexplored cause of transient
execution.
We show that invoking \emph{microcode assists} to handle corner cases
in the execution of some instructions, results in transient
execution of the instructions.
While assists-based transient execution shares some properties of
Meltdown-type transient execution,
assists do not cause exceptions and therefore do not require any
fault-suppression mechanisms.

\subsection{Disclosure and Timeline}
\noindent Following the practice of responsible disclosure, we have notified
CPU vendors about our findings. 

\parhead{Intel.} We notified Intel about our findings, including a
preliminary writeup and proof-of-concept code, on January 31st, 2019. Intel had
acknowledged the issue and requested an
embargo on the results in this paper, ending May 14th, 2019. Intel has further classified this issue as 
Microarchitectural Store Buffer Data Sampling (MSBDS), assigning it \texttt{CVE-2018-12126} and a CVSS ranking of Medium.
Finally, Intel had indicated that
we are the first academic group to report this issue and  that a similar issue
was found internally as well.

\parhead{AMD.} We also notified AMD's security response team regarding our
findings, including our writeup. AMD had investigated this issue of their
architectures and indicated that AMD CPUs are not vulnerable to the attacks
described in this paper.

\parhead{ARM.} We have also notified ARM's security response team regarding our findings.
ARM had investigated this issue and found that ARM CPUs are not vulnerable to the attacks described in this paper.

\parhead{IBM.} Finally, we also notified IBM security about the finding
reported in this work. IBM had responded that none of their CPUs is affected,
including System-V and PowerPC.

\parhead{The RIDL Attack.}
In a concurrent independent work\footnote{Both teams made contact on May 7th,  provided each other with an overview of their findings, and coordinated public disclosure as well as communication with Intel. For a complete timeline describing the flow of information related to this disclosure, see \url{mdsattacks.com}.}, the RIDL attack~\cite{vanSchaik2019ridl} 
analyzes additional buffers present inside Intel CPUs, with specific attention
to the Line Fill Buffer (LFB) and load ports. There, they show that faulty loads from the LFB or load ports 
leak information across various security domains.  We note however that
\AttackName is different from (and complementary to) RIDL. This is since the
two attacks exploit different microarchitectural elements (LFB and load ports for RIDL and
Store Buffer and WTF optimization for \AttackName).  In particular, RIDL can be
used to recover values recently placed in the LFB while \AttackName allows the
attacker to recover the value of a specific attacker-chosen writes in the store buffer.

\section{Background}
In this section, we provide the background required to understand our attack,
including a description of caches and cache attacks, transient execution attacks,
and Intel Transactional Synchronization Extensions.

\subsection{Caches and Cache Attacks}
Caches are an essential part of modern processors.  They are small and fast
memories where the CPU stores copies of data from the main memory to hide the
main memory access latency.  Modern CPUs have a variety of different caches and
buffers for various purposes.  The main cache hierarchy is the instruction and
data cache hierarchy consisting of multiple levels, which vary in size and
latency.  The L1 is the smallest and fastest cache.  The L3 cache, also called
the last-level cache (LLC), is typically the largest and slowest.

\parhead{Cache Organization.}
Modern caches are typically set-associative, \ie a cache line is stored in a
fixed set, as determined by part of its virtual or physical address.  Addresses
that map to the same set are called \emph{congruent}.  On modern processors, the
last-level cache is typically physically indexed and shared across cores.  It
is also often inclusive of L1 and L2, which means that all data stored in L1
and L2 is also stored in the last-level cache.  The cache hierarchy exposes the
latency difference between the main memory access (cache miss) and the cache
access (cache hit), \ie exactly the latency difference that caches introduce.
This can be used in side channels on a non-colluding victim or in covert
channels where sender and receiver collude to transmit information.

\parhead{Cache Attacks.}
Different cache attack techniques have been proposed in the past, such as
\PrimeProbe~\cite{Osvik2006,Percival2005} and
\FlushReload~\cite{yarom2014flush+}.  \FlushReload attacks and its
variants~\cite{Gruss2015Template,Lipp2016,Gruss2016Flush,ZhangXZ16} work on shared memory
at a cache-line granularity.  The attacker repeatedly flushes a cache line
and measures how long it takes to reload it.  The reload time will always be
high unless another process has reloaded the cache line back into
the cache.  In contrast, \PrimeProbe attacks work without shared memory, and only at a
cache-set granularity.  The attacker repeatedly accesses a set of congruent
memory addresses, filling an entire cache set with its own cache lines, and
measures how long that takes.  As this is repeated in a loop, the cache set 
is always filled with the attacker's cache lines. Hence the access time
will always be rather low. However, if another process accesses a memory
location in the same cache set, it will evict one of the attacker's cache lines
and the access time will increase.

Cache attacks have been used to break cryptographic
implementations~\cite{Osvik2006,Percival2005,Liu2015,yarom2014flush+,genkin2017may,
YaromGH17, genkin2018drive}, infer user
input~\cite{Gruss2015Template,Lipp2016,Schwarz2018KeyDrown}, and break
system-level security~\cite{Hund2013,Gruss2016Prefetch}.  Both \PrimeProbe and
\FlushReload have also been used in high-performance covert
channels~\cite{Liu2015,Maurice2017Hello,Gruss2016Flush}, also as a building
block of transient execution attacks such as Meltdown~\cite{Lipp2018meltdown},
Spectre~\cite{Kocher2019spectre}, and Foreshadow~\cite{Vanbulck2018foreshadow,
weisse2018foreshadowNG} that we detail below.

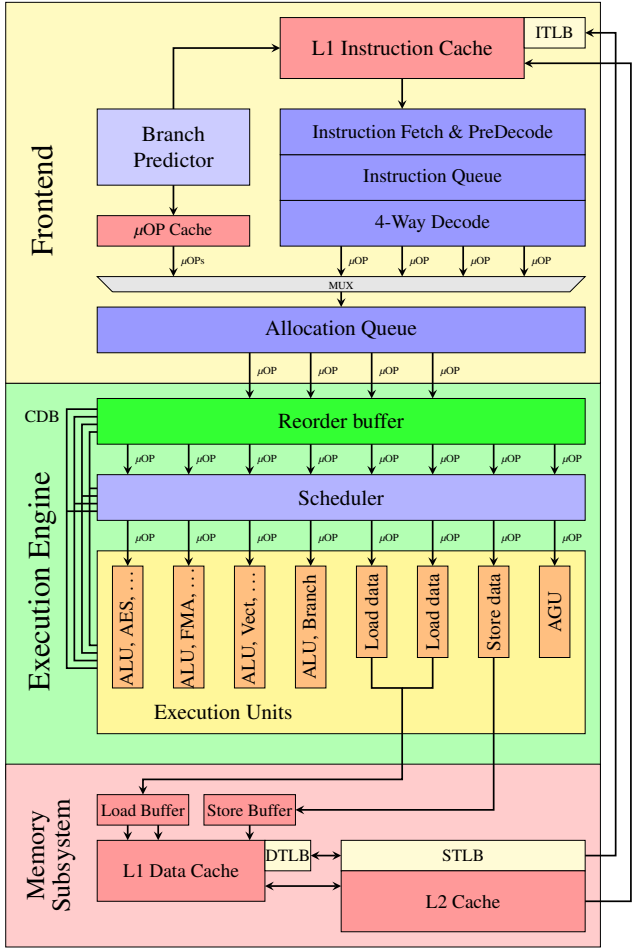
\begin{figure}[t!]
	\centering
	\resizebox{\hsize}{!}{%
		\begin{tikzpicture}

\tikzstyle{execution_engine}+=[fill=green!30];
\tikzstyle{reorder_buffer}+=[fill=green!80];
\tikzstyle{scheduler}+=[fill=blue!30];
\tikzstyle{execution_units}+=[fill=yellow!50];
\tikzstyle{execution_unit}+=[fill=orange!50];

\tikzstyle{memory_subsystem}+=[fill=red!20];
\tikzstyle{tlb}+=[fill=yellow!20];
\tikzstyle{cache}+=[fill=red!40];

\tikzstyle{frontend}+=[fill=yellow!30];
\tikzstyle{bpu}+=[fill=blue!20];
\tikzstyle{instruction_fetch}+=[fill=blue!40];
\tikzstyle{instruction_queue}+=[fill=blue!40];
\tikzstyle{waydecode}+=[fill=blue!40];
\tikzstyle{mux}+=[fill=black!10];
\tikzstyle{allocation_queue}+=[fill=blue!40];

\tikzstyle{myarrow} += [->,>=stealth,thick];

\draw[execution_engine] (-1.5,0.75) rectangle +(9.75,-6.5);
\draw (-0.9,-2.5) node [rotate=90] {\Large Execution Engine};

\draw [reorder_buffer] (0,-0.25) rectangle +(8, 0.75) node [pos=.5] {Reorder buffer};
\draw [myarrow] (0.5,-0.25) -- +(0,-0.5) node [pos=.5,right] {\tiny $\mu$OP};
\draw [myarrow] (1.5,-0.25) -- +(0,-0.5) node [pos=.5,right] {\tiny $\mu$OP};
\draw [myarrow] (2.5,-0.25) -- +(0,-0.5) node [pos=.5,right] {\tiny $\mu$OP};
\draw [myarrow] (3.5,-0.25) -- +(0,-0.5) node [pos=.5,right] {\tiny $\mu$OP};
\draw [myarrow] (4.5,-0.25) -- +(0,-0.5) node [pos=.5,right] {\tiny $\mu$OP};
\draw [myarrow] (5.5,-0.25) -- +(0,-0.5) node [pos=.5,right] {\tiny $\mu$OP};
\draw [myarrow] (6.5,-0.25) -- +(0,-0.5) node [pos=.5,right] {\tiny $\mu$OP};
\draw [myarrow] (7.5,-0.25) -- +(0,-0.5) node [pos=.5,right] {\tiny $\mu$OP};

\draw [scheduler] (0,-0.75) rectangle +(8, -0.75) node [pos=.5] {Scheduler};

\draw [execution_units] (0,-2) rectangle +(8, -3) node [yshift=-32.5,xshift=-55,pos=.5] {Execution Units};
\draw [execution_unit] (0.25, -2.25) rectangle +(0.5,-2) node [rotate=90,pos=.5] {\small ALU, AES, \dots};
\draw [execution_unit] (1.25, -2.25) rectangle +(0.5,-2) node [rotate=90,pos=.5] {\small ALU, FMA, \dots};
\draw [execution_unit] (2.25, -2.25) rectangle +(0.5,-2) node [rotate=90,pos=.5] {\small ALU, Vect, \dots};
\draw [execution_unit] (3.25, -2.25) rectangle +(0.5,-2) node [rotate=90,pos=.5] {\small ALU, Branch};
\draw [execution_unit] (4.25, -2.25) rectangle +(0.5,-1.5) node [rotate=90,pos=.5] {\small Load data};
\draw [execution_unit] (5.25, -2.25) rectangle +(0.5,-1.5) node [rotate=90,pos=.5] {\small Load data};
\draw [execution_unit] (6.25, -2.25) rectangle +(0.5,-1.5) node [rotate=90,pos=.5] {\small Store data};
\draw [execution_unit] (7.25, -2.25) rectangle +(0.5,-1.5) node [rotate=90,pos=.5] {\small AGU};

\draw [myarrow] (0.5,-1.5) -- +(0,-0.75) node [pos=.4,right] {\tiny $\mu$OP};
\draw [myarrow] (1.5,-1.5) -- +(0,-0.75) node [pos=.4,right] {\tiny $\mu$OP};
\draw [myarrow] (2.5,-1.5) -- +(0,-0.75) node [pos=.4,right] {\tiny $\mu$OP};
\draw [myarrow] (3.5,-1.5) -- +(0,-0.75) node [pos=.4,right] {\tiny $\mu$OP};
\draw [myarrow] (4.5,-1.5) -- +(0,-0.75) node [pos=.4,right] {\tiny $\mu$OP};
\draw [myarrow] (5.5,-1.5) -- +(0,-0.75) node [pos=.4,right] {\tiny $\mu$OP};
\draw [myarrow] (6.5,-1.5) -- +(0,-0.75) node [pos=.4,right] {\tiny $\mu$OP};
\draw [myarrow] (7.5,-1.5) -- +(0,-0.75) node [pos=.4,right] {\tiny $\mu$OP};

\draw [myarrow,-] (0, -3.55) -- ++(-0.125,0) -- ++(0,0.1+3.75-0.35) -- ++(0.125,0);
\draw [myarrow,-] (0, -3.55-0.125) -- ++(-0.25,0) -- ++(0,0.1+4-0.35) -- ++(0.25,0);
\draw [myarrow,-] (0, -3.55-0.25) -- ++(-0.375,0) -- ++(0,0.1+4.25-0.35) -- ++(0.375,0);
\draw [myarrow,-] (0, -3.55-0.375) -- ++(-0.5,0) -- ++(0,0.1+4.5-0.35) -- ++(0.5,0);

\draw [myarrow,-] (0, -0.875-0.1) -- ++(-0.125,0);
\draw [myarrow,-] (0, -1-0.1) -- ++(-0.25,0);
\draw [myarrow,-] (0, -1.125-0.1) -- ++(-0.375,0);
\draw [myarrow,-] (0, -1.25-0.1) -- ++(-0.5,0);

\draw (-0.9,0.2) node {\footnotesize CDB};

\draw [memory_subsystem] (-1.5,-5.5) rectangle +(9.75, -3);
\draw (-0.75,-7) node [rotate=90, text width=2cm, align=center] {\large Memory\\ Subsystem};

\draw [cache] (0, -6) rectangle +(1.5,-0.5) node [pos=.5] {\footnotesize Load Buffer};
\draw [myarrow,<-] (0.75,-6) -- ++(0, 0.25) -- ++(4.25,0) |- (4.5, -4.25) -- ++(0,0.5);
\draw [myarrow,-] (5,-4.25) |- (5.5, -4.25) -- +(0,0.5);

\draw [myarrow] (0.5,-6.5) -- ++(0,-0.25);
\draw [myarrow] (1,-6.5) -- ++(0,-0.25);

\draw [cache] (1.75, -6) rectangle +(1.5,-0.5) node [pos=.5] {\footnotesize Store Buffer};
\draw [myarrow] (6.5, -3.75) -- ++(0,-2.5) -- ++(-3.25,0);

\draw [myarrow] (2.5,-6.5) -- ++(0,-0.25);

\draw [cache] (0, -6.75) rectangle +(2.75,-1) node [pos=.5] {\small L1 Data Cache};
\draw [tlb] (2.75, -6.75) rectangle +(0.75,-0.5) node [pos=.5] {\footnotesize DTLB};

\draw [myarrow,<->] (3.5,-7) -- ++(0.5,0);
\draw [myarrow,<->] (2.75,-7.5) -- ++(1.25,0);

\draw [tlb] (4,-6.75) rectangle ++(4,-0.5) node [pos=.5] {\footnotesize STLB};
\draw [cache] (4,-7.25) rectangle ++(4,-1) node [pos=.5] {\small L2 Cache};

\begin{scope}[yshift=-0.25cm]
\draw [frontend] (-1.5,1) rectangle +(9.75,6.25);
\draw (-0.9,4) node [rotate=90] {\Large Frontend};

\draw [allocation_queue](0,1.5) rectangle ++(8,0.75) node [pos=.5] {Allocation Queue};
\draw [myarrow] (2.5,1.5) -- +(0,-0.75) node [pos=.4,right] {\tiny $\mu$OP};
\draw [myarrow] (3.5,1.5) -- +(0,-0.75) node [pos=.4,right] {\tiny $\mu$OP};
\draw [myarrow] (4.5,1.5) -- +(0,-0.75) node [pos=.4,right] {\tiny $\mu$OP};
\draw [myarrow] (5.5,1.5) -- +(0,-0.75) node [pos=.4,right] {\tiny $\mu$OP};

\draw [-,mux] (0.25,2.5) -- ++(7.5,0) -- ++(0.25,0.25) -- ++(-8,0) -- ++(0.25,-0.25);
\draw (4,2.5) node [yshift=3] {\tiny MUX};
\draw [myarrow] (4,2.5) -- ++(0,-0.25);

\draw [waydecode] (3,3.25) rectangle ++(5,0.75) node [pos=.5] {\small 4-Way Decode};
\draw [myarrow] (4,3.25) -- +(0,-0.5) node [pos=.5,right] {\tiny $\mu$OP};
\draw [myarrow] (5,3.25) -- +(0,-0.5) node [pos=.5,right] {\tiny $\mu$OP};
\draw [myarrow] (6,3.25) -- +(0,-0.5) node [pos=.5,right] {\tiny $\mu$OP};
\draw [myarrow] (7,3.25) -- +(0,-0.5) node [pos=.5,right] {\tiny $\mu$OP};

\draw [instruction_queue] (3,4) rectangle ++(5,0.75) node [pos=.5] {\small Instruction Queue};

\draw [instruction_fetch] (3,4.75) rectangle ++(5,0.75) node [pos=.5] {\small Instruction Fetch \& PreDecode};

\draw [cache] (0,3.25) rectangle ++(2.5,0.5) node [pos=.5] {\footnotesize $\mu$OP Cache};
\draw [myarrow] (1.25,3.25) -- ++(0,-0.5) node [pos=.5,right] {\tiny $\mu$OPs};

\draw [bpu] (0,4.25) rectangle ++(2.5,1.25) node [pos=.5, text width=2cm, align=center] {Branch\\ Predictor};
\draw [myarrow] (1.25,4.25) -- ++(0,-0.5);

\draw [cache] (3,6) rectangle ++(4,1) node [pos=.5] {L1 Instruction Cache};
\draw [tlb] (7,6.5) rectangle ++(1,0.5) node [pos=.5] {\footnotesize ITLB};

\draw [myarrow,<-] (3,6.5) -- ++(-1.75,0) -- ++(-0,-1);
\draw [myarrow] (5,6) -- ++(0,-0.5);
\end{scope}

\draw [myarrow] (8,-7.75) -- ++(0.75,0) -- ++(0,13.75) -- ++(-1.75,0);
\draw [myarrow] (8,-7) -- ++(0.5,0) -- ++(0,13.5) -- ++(-0.5,0);

\end{tikzpicture}
	}%
	\caption{Simplified illustration of a single core of the Intel's Skylake microarchitecture (as presented in~\cite{Lipp2018meltdown}). Instructions are decoded into \muops and executed out-of-order in the execution engine by individual execution units.}
	\label{fig:core-skylake}
\end{figure}

\subsection{Superscalar Processors}
To achieve their high performance, modern processors are often \emph{superscalar},
that is, they perform multiple operations in parallel.
In current implementations, \eg in modern Intel processors (refer \cref{fig:core-skylake}),
execution of a program is divided between two main parts.
The \emph{frontend} is responsible for processing the machine-code instructions
of the program, decoding them to a stream of \emph{micro-ops} (\muops) that are
sent to the \emph{Execution Engine} for execution.

\parhead{Out-of-order Execution.}
The execution engine consists of multiple execution units, which can execute various
\muops.
To allow superscalar execution, the execution engine follows a variant of 
Tomasulo's algorithm~\cite{Tomasulo1967}, which executes \muops when the data
they depend on is available, rather than following strict program order.
Once executed, the \muops arrive at the \emph{reorder buffer} whose
purpose is to \emph{retire} \muops in program order,
ensuring that architecturally-visible effects of \muops execute in the
order the programmer specified.

\parhead{Speculative Execution.}
The stream of \muops that the frontend generates does not necessarily correspond 
to the sequence of instructions in the program.
A major cause of deviation is \emph{branch prediction}.
When the frontend reaches a branch instruction, it often does not
yet know where execution will proceed.
Instead of waiting, the frontend attempts to predict the outcome
of the branch and proceed from there.
In the case that the prediction is correct, the generated \muops
match the program and can be processed.
Otherwise, at some later stage, the processor notices the \emph{misprediction}.
The frontend is then \emph{steered} to the correct instruction, and
\muops generated as part of the misprediction are dropped by the
reorder buffer without committing any of their results to the architectural
state of the processor.
Following \cite{canella2018systematic}, we refer to \muops that are
not retired as \emph{transient}.
Similarly, following \cite{glew1997method}, we use refer to \muops other
than the one waiting for retirement as \emph{speculative}.
We note that speculative \muops do not necessarily result
from speculative execution.  The are called speculative because
the execution engine cannot determine whether they are transient or not.

\subsection{The Memory Subsystem}
In this work, we are mainly interested in how memory load and store operations
are implemented.
The main two issues we deal with are how to resolve the physical
addresses used by these instructions and how to ensure that out-of-order
execution does not break dependencies between these instructions.

\subsection{Transient Execution Attacks}\label{sec:speculative-attacks}

While transient execution does not influence the architectural state of the processor, it can
change the microarchitectural state.  Transient execution attacks abuse
transient execution to execute a few instructions transiently and 
modify the microarchitectural state.  The change in the microarchitectural
state is then observed using a covert-channel attack.
Spectre-type~\cite{Kocher2019spectre} attacks exploit different prediction
mechanisms, while Meltdown-type~\cite{Lipp2018meltdown,Vanbulck2018foreshadow}
attacks exploit transient execution following a CPU exception.

\parhead{Spectre Attacks.}
The first Spectre attacks focused on the CPU's Pattern History Table (PHT),
Branch History Buffer (BHB), and Branch Target Buffer (BTB)  as
microarchitectural data structures causing
mispredictions~\cite{Kocher2019spectre}.  Both transient loads and
stores~\cite{Kiriansky2018speculative} are possible, leading to a variety of
attacks, including reading and writing from out-of-bound memory locations,
transferring control-flow to arbitrary addresses via mispredicted indirect
jumps~\cite{Kocher2019spectre} or returns~\cite{Koruyeh2018spectre5,
Maisuradze2018spectre5}.  In all Spectre attacks, the attacker mistrains the
processor by performing a certain type of branches, influencing the
corresponding microarchitectural predictor.  Subsequently, the victim runs with
incorrect predictions and thereby leaks data.  While Spectre attacks can only leak architecturally accessible data, the mistraining works across privilege boundaries, \eg the kernel-to-user boundary, or SGX.  Another type of
Spectre attacks is based on unsuccessful load-to-store
forwarding~\cite{Horn2018spectre4}.  Spectre attacks can even be mounted in
remote scenarios, \ie from JavaScript~\cite{Kocher2019spectre} or just by
sending requests to a vulnerable system~\cite{Schwarz2018netspectre}.

\parhead{Meltdown Attacks.}
Meltdown-type attacks do not exploit misprediction.  Instead, they exploit
deferred handling of permission checks.  Before the permission check is
performed and the attacker process triggers a processor exception
architecturally, the data is already handed to the subsequent instructions that
are also transiently executed.  The first Meltdown
attack~\cite{Lipp2018meltdown} exploits the deferred permission check for the
user/supervisor bit in the page tables, allowing to leak arbitrary memory
mapped in the kernel address space.  Other Meltdown attacks similarly exploit the deferred check
of present or reserved bits in page table
entries~\cite{Vanbulck2018foreshadow,weisse2018foreshadowNG}, the writable
bit in the page table entry~\cite{Kiriansky2018speculative}, or the permission check when reading
system registers~\cite{ARMSpecAnalysis,IntelSpecAnalysis}.

\parhead{Countermeasures.}
Recognizing the danger posed by transient execution attacks, a wide range of
defenses have been proposed to defend against them.  However, to date, it is
unclear which defenses actually increase the security level and which are
trivially bypassable~\cite{Mcilroy2019spectre,canella2018systematic}.  One
defense where the consensus across academia and industry is that it protects against Meltdown, if
correctly implemented, is KAISER~\cite{Gruss2017KASLR}.  KAISER is the idea of
duplicating the page table hierarchies for every process, once with the kernel
space mappings present and once without.  When running in user space the
mapping without the kernel space is used.  The idea of KAISER has been
integrated into all major operating systems, \eg in Linux as
KPTI~\cite{LWN_kpti}, in Windows as KVA Shadow~\cite{Ionescu2017Twitter}, and
in Apple's xnu kernel as double map~\cite{Levin2012}.  While KAISER costs
performance, the use of PCID and ASID on modern processors reduced the
overheads for real-world workloads to almost zero~\cite{Gregg2018kpti}.  More
recent processors ship with hardware patches and hence have the KAISER patch
disabled by default~\cite{i9eval}.

\subsection{Exception Creation} 
As explained in \cref{sec:speculative-attacks}, in a Meltdown-type
attack the attacker exploits the deferred enforcement of permissions (i.e.,
deferred exception handling) present in Intel CPUs in order to obtain
privileged information.   In the original Meltdown
attack~\cite{Lipp2018meltdown}, the attacker exploits the delayed enforcement
of the User / Supervisor bit in the CPU's hardware in order to read privileged
information and subsequently leak it through a covert channel. Next, in
Foreshadow~\cite{Vanbulck2018foreshadow} and
Foreshadow-NG~\cite{weisse2018foreshadowNG}, the attacker exploits the fault
cases of a page marked as non-present and therefore cannot be accessed.

\subsection{Exception Suppression}\label{sec:exception_suppression}
One problem common to Meltdown-type attacks is that the instructions they
exploit cause exceptions, which by default terminate the program.  Four main
approaches have been suggested for handling this termination.  In the
fork-and-crash approach, a forked process executes the attack, and its parent
resumes after the process terminates. Exception handling sets up a signal
handler to catch the exception and resume execution.  A third option suppresses
the exception by wrapping the attack code in a mispredicted branch or call, which
speculatively executes the attack.  Finally, the exception can be suppressed by
wrapping it in a hardware transaction.  The last approach is the most
effective~\cite{Lipp2018meltdown} and most widely
applicable~\cite{Vanbulck2018foreshadow, weisse2018foreshadowNG}. Given its
applicability, in \cref{sec:tsx} below, we provide additional details
about exception suppressing using hardware transactions. We refer interested
readers to \cite{Lipp2018meltdown} for further information on the other
approaches.

\subsection{Transactional Memory}\label{sec:tsx}
Intel's Transactional Synchronization Extensions (TSX) is an instruction set
extension to the x86-64 architecture that supports hardware transactions.  In a
nutshell, a transaction is a sequence of instructions that are either executed
atomically or not executed at all.  Atomic execution implies that concurrent
threads cannot observe intermediate updates from the transaction and the thread
executing the transaction cannot observe any changes from other threads.

\parhead{Implementing TSX Transactions.}
Transactions are delimited by two instructions. 
The \instr{xbegin} instruction starts a transaction and \instr{xend} terminates
it. The \instr{xbegin} instruction also specifies an abort location where
execution continues if the transaction fails. 
Transaction implementation mostly relies on existing processor mechanisms.
Instructions following \instr{xbegin} are not retired and instead are kept in the
reorder buffer until the \instr{xend} is executed. 
If the transaction is aborted, all pending instructions
in the transaction are discarded, and the architectural state of the processor is reverted
to the state before the \instr{xbegin}.
To revert memory state and to maintain atomicity, memory stores inside a transaction
modify the L1 cache but are not evicted to lower memory layers, and memory
lines read in a transaction remain in the last-level cache.
TSX locks the affected lines to protect against concurrent modifications and reads of modified lines.

\parhead{Transaction Aborts.} 
If concurrent processes try to write to these locked lines, the transaction
aborts and is rolled back.
 Similarly, if the
processor runs out of cache space for the transaction data, the transaction
aborts.  This behavior of TSX transactions has been exploited for both
side-channel attacks and
defenses~\cite{DisselkoenKPT17,Shih0KP17,GrussLSOHC17}.  Transactions also
abort in other scenarios.  In particular, transactions abort when the processor
receives an exception or if an instruction within the transaction causes a
fault.  Thus, when a Meltdown-type attack is enclosed in a TSX transaction, the
faulting instruction causes a transaction abort, which effectively reverses the
architectural state of the processor to the state prior the \instr{xbegin}
instruction, suppressing the fault.  Yet, as \mbox{\cite{Lipp2018meltdown}} observe,
the microarchitectural state of the processor is not reverted when a
transaction aborts, allowing the attacker to recover information from the
aborted instructions.

\section{The Write Transient Forwarding Optimization}
In this section, we discuss the WTF optimization that is exploited with the \AttackName attack.
First, we will illustrate the basic idea of \AttackName with a simple toy example before discussing the hardware mechanisms responsible for the attack.

\subsection{A Toy Example}\label{sec:toy-read}
\Cref{lst:pseudocode} shows a simple code snippet which exploits the WTF optimization to read variables without directly accessing them.
While this example does not have security implications on its own, it nonetheless shows the general concept behind \AttackName, allowing user-level code to read information stored in the CPU's store buffer without directly accessing the address corresponding to that information.

\begin{lstlisting}[language=C,style=customc, float=htb, label={lst:pseudocode}, caption={Pseudocode of \AttackName. Some mmap parameters were omitted for clarity},escapechar=|]
char* victim_page = mmap(..., PAGE_SIZE, ...);|\label{line:mmap-target}|
char* attacker_page = mmap(..., PAGE_SIZE, ...);|\label{line:mmap-attacker}|
mprotect(attacker_page, PAGE_SIZE, PROT_NONE);|\label{line:mprotect-attacker}|

offset = 7; |\label{line:target-address}|
victim_page[offset] = 42; |\label{line:target-write}|

if (tsx_begin() == 0) {|\label{line:tsx}|
  memory_access(lut + 4096 * attacker_page[offset]); |\label{line:spec-trans}|
  tsx_end();
}

for (i = 0; i < 256; i++) {|\label{line:probe-start}|
  if (flush_reload(lut + i * 4096)) { |\label{line:reload}|
    report(i);
  }
}|\label{line:probe-end}|
\end{lstlisting}

\parhead{Setup.} First, 2 pages are allocated.
The \texttt{victim\_page} is a user space accessible page where the user can store and read data.
However, by setting the protection level to \texttt{PROT\_NONE} on the \texttt{attacker\_page}, all access permissions to this page are revoked and the page is marked as \textit{not-present}.
Thus, any access to the \texttt{attacker\_page} will yield an exception.

Next, we write the value 42 to the offset 7 of the \texttt{victim\_page}.
Rather than executing the write to memory immediately, the processor first notes the operation in the store buffer.
We note that the code in \cref{lst:pseudocode} never reads from the \texttt{victim\_page} directly.

\parhead{Reading Previous Stores.}
Instead of reading from the victim page at the specified offset, the code starts a TSX transaction (\cref{line:tsx}) and reads from the \texttt{attacker\_page}.
As the page is inaccessible, the memory access will fail and the TSX transaction aborts.
However, the exception will be only handled by the reorder buffer when the memory access operation is retired.
In the meantime, due to the WTF optimization, the CPU will transiently forward the value of the previous store at the same page offset.
Thus, the memory access will pick-up the value of the store to the \texttt{victim\_page}, in this example 42.
Using a cache-based covert channel, the incorrectly forwarded value is transmitted.
Finally, when the failure and transaction abort are handled, the architectural effects of the transiently executed code are reverted.

\begin{figure}[t!]
  \centering
  \includegraphics[width=\linewidth]{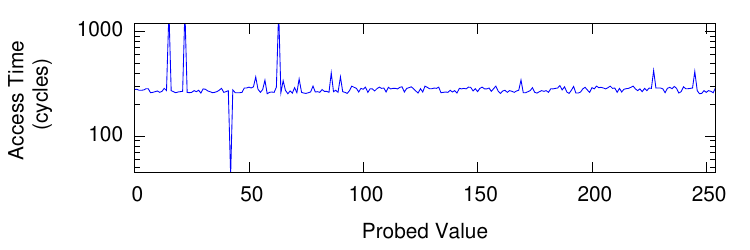}
  \caption{Access times to the probing array during the execution of
    \cref{lst:pseudocode}. The dip at 42
    represents a correct recovery of the value from the store buffer. }
  \label{fig:probe-toy}
\end{figure}

\parhead{Recovering the Leaked Data.}
Using Flush+Reload, the attacker can recover the leaked value from the cache-based covert channel in \cref{line:reload}.
\cref{fig:probe-toy} displays the results of measured access times to the look-up-table (\texttt{lut}) on a Meltdown-resistant i9-9900K CPU.
As the figure illustrates, the typical access time to an array element is above 200 cycles, with the exception of element 42, where the access time is well below 100 cycles.
We note that this position matches the value written to \texttt{target\_page}.
Hence, the code can recover the value without directly reading it.

\subsection{The Mechanism Behind \AttackName}\label{sec:fallout-mech}
We now turn our attention to the \emph{store buffer}, a microarchitectural
component, which lies in the core of WTF and \AttackName.

\parhead{The Store Buffer Implementation.}
When the CPU writes data to memory, it needs to first resolve the virtual address
to a physical address. 
Then it acquires exclusive access to the cache line of the target data. 
Rather than waiting, the processor stores the information to the store buffer.

\begin{figure}[t!]
	\centering
	\vspace{-2em}
	\includegraphics[width=\linewidth]{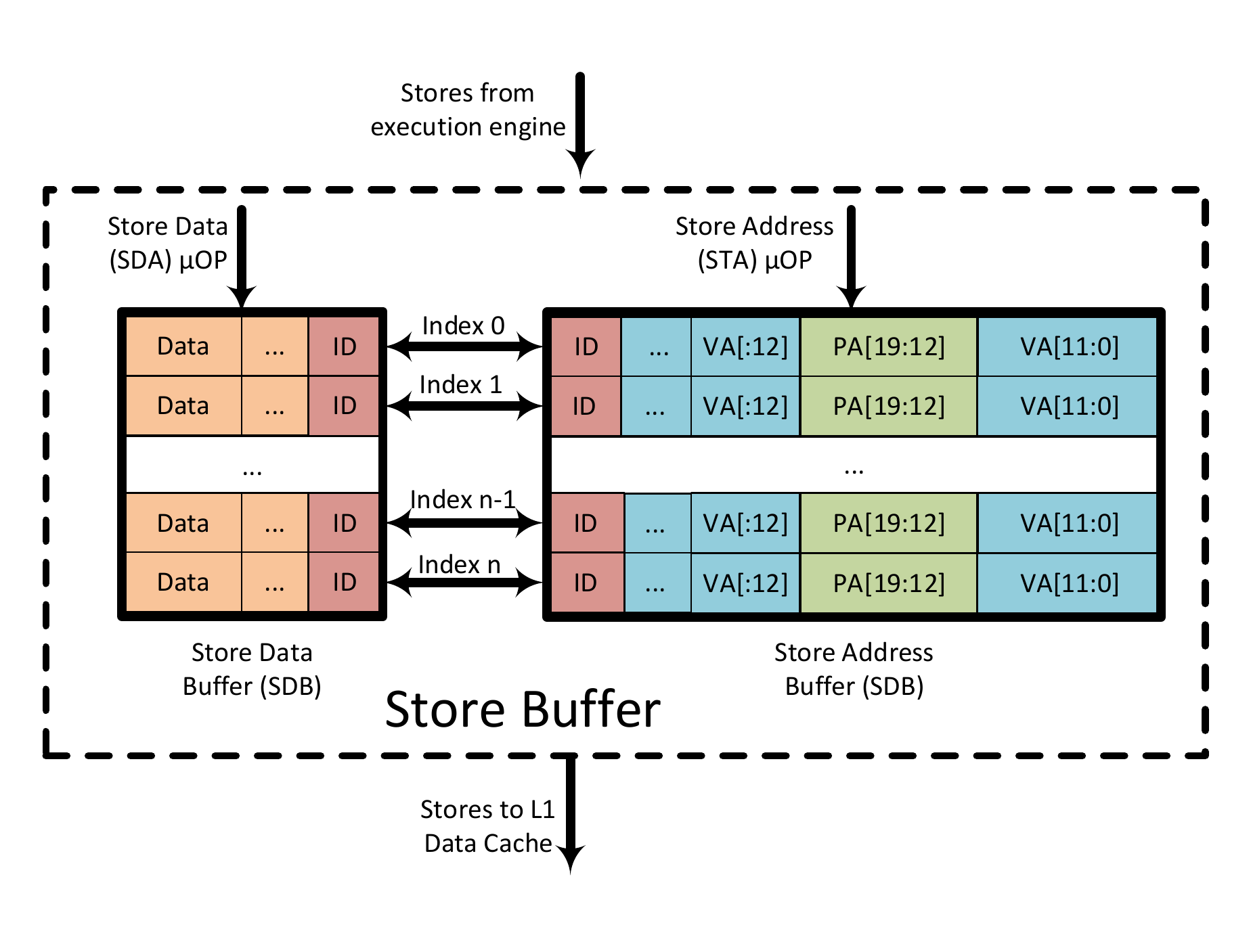}
	\vspace{-3.5em}
	\caption{Structure of the store buffer on Intel CPUs.}
	\label{fig:write-buffer}
\end{figure}

\cref{fig:write-buffer} shows the structure of the store buffer according
to Intel patents~\cite{abramson1998method, abramson2002method}.  
Based on these patents, a store operation is implemented using two \muops,
store address (STA) and store data (SDA).
Splitting the operation to two \muops allows the processor to process the
parts independently and asynchronously.

Asynchronous processing raises the issue of memory ordering.
Specifically, operations that access the same memory locations must
be performed at the order specified in the program and,
in particular, load operations should get the value from preceding 
stores to the same address.
Intel published some properties of the store buffer~\cite{Intel_64_IA-32:aorm}.
However, we are not aware of any public documentation of the algorithms
used for resolving memory access conflicts. 
Intel's patents on the 
topic~\cite{abramson1998method, abramson2002method, kosinski2012store}
suggest
that the store buffer is virtually indexed, but each entry also includes parts
of the physical address, such that mismatches on the partial addresses ensure
the absence of dependencies, allowing loads to proceed without waiting for full
address resolution.

\parhead{Write Transient Forwarding.}
An algorithm for handling partial address matches appears in another
Intel patent~\cite{HilyZH09}.
Remarkably, the patent explicitly states that:
\begin{quote}
  "if there is a hit at operation 302 [partial match using  page offsets] 
  and the physical address of the load or the store operations is
  not valid, the physical address check at operation 310 [full physical address match]
  may be considered as a hit"
\end{quote}
That is, if address translation of a load \muop fails and the 12 least
significant bits of the load address match those of a prior store,
the processor assumes that the physical addresses of the load and the store
match and forwards the previously stored value to the load \muop.
We note that the failed load is transient and will not retire, hence WTF has
no architectural implications.  
However, as this work demonstrates, microarchitectural side effects of
transient execution following the failed load may result in inadvertent
information leaks.
Given the surprising nature of this optimization and its security
consequence, we refer to it as the Write Transient Forwarding (WTF) optimization.

\newlength{\precl}
\settowidth{\precl}{Pre CL}

\begin{table*}[t!]
	\centering
	\begin{tabular}{lcccccc} \hline
		\textbf{Fault Suppression} & \multicolumn{2}{c}{\bf Transactional Memory (TSX)}               & \multicolumn{2}{c}{\bf Branch Misprediction}\\
		Architecture           & \cellcolor{tabledark}{Pre Coffee Lake R} &{Coffee Lake R} & \cellcolor{tabledark}{Pre Coffee Lake R} &{Coffee Lake R} \\ \hline
		User not present  & \cellcolor{tabledark}\xmark & \ccmark & \cellcolor{tabledark}\xmark & \xmark  \\
		Kernel data       & \cellcolor{tabledark}\cmark & \cxmark & \cellcolor{tabledark}\cmark & \cxmark \\
		Kernel code       & \cellcolor{tabledark}\cmark & \cmark  & \cellcolor{tabledark}\cmark & \cmark  \\
		Kernel not present & \cellcolor{tabledark}\xmark & \xmark  & \cellcolor{tabledark}\xmark & \xmark \\
		SMAP              & \cellcolor{tabledark}\cmark & \cmark  & \cellcolor{tabledark}\cmark & \cxmark \\
		\hline
	\end{tabular}
	\caption{Evaluating different fault-inducing and fault-suppression mechanisms
		on Intel architectures before Coffee Lake R and on Coffee Lake R. 
		\cmark indicates that our attack can successfully leak data, while \xmark indicates no leakage was
		observed.  Finally, we denote the case of the Coffee Lake R regression with
		\textccmark, while changes following hardware countermeasures are marked with
		\textcxmark.  
	}
	\label{table:sbr-table}
\end{table*}

\parhead{Fault and Suppression Mechanisms.}
To better understand the WTF mechanism, we evaluate the toy example in
\cref{lst:pseudocode} with multiple combinations of causes of faults and 
fault-suppression mechanisms.
We experimented with three Intel processors: 
a Coffee Lake R i9-9900K, a Kaby Lake i7-7600U, and a Skylake i7-6700.
We summarize the results in \cref{table:sbr-table}.

We observe that unlike earlier generations, the Coffee Lake R processor 
exhibits a different behavior based on the fault suppression mechanism.
Specifically, in the example in \cref{lst:pseudocode} replacing the TSX
fault suppression mechanism with branch misprediction does not trigger the
WTF optimization, and the value does not leak.
We suspect that the processor inhibits some forms of speculative
execution within branch misprediction while allowing it in TSX
transactions.
Moreover, the Coffee Lake R processor does not seem to trigger the WTF optimization
when a load fails due to a read from a kernel page.
We note that transient reads from such pages is the main cause of the
Meltdown bug.
Thus, we conjecture that the differences in behavior between
the processor generations are
due to the recent mitigations for the Meltdown and Foreshadow attacks 
introduced in the Coffee Lake R architecture~\refs. 

Also note that, for completeness, we tested whether WTF can be triggered by
Supervisor Mode Access Prevention (SMAP) features in recent Intel processors.
For this experiment, we explicitly dereference a user space pointer in kernel
mode such that SMAP raises an architectural fault.
We observed that SMAP violations may successfully trigger the WTF optimization.
While we do not consider this to be an exploitable attack scenario, SMAP was to te
best of our knowledge previously considered to be immune to any Meltdown-type
effects~\cite{canella2018systematic}.

\parhead{Coffee Lake R Regression.} We also note a troubling \emph{regression} in
Intel's newest architecture.
When accessing a page marked as non-present, we can only trigger the
WTF optimization on the Coffee Lake Refresh processor.



\subsection{Measuring the Store Buffer Size}

\begin{figure}[t!]
	\centering
	\includegraphics[width=\linewidth]{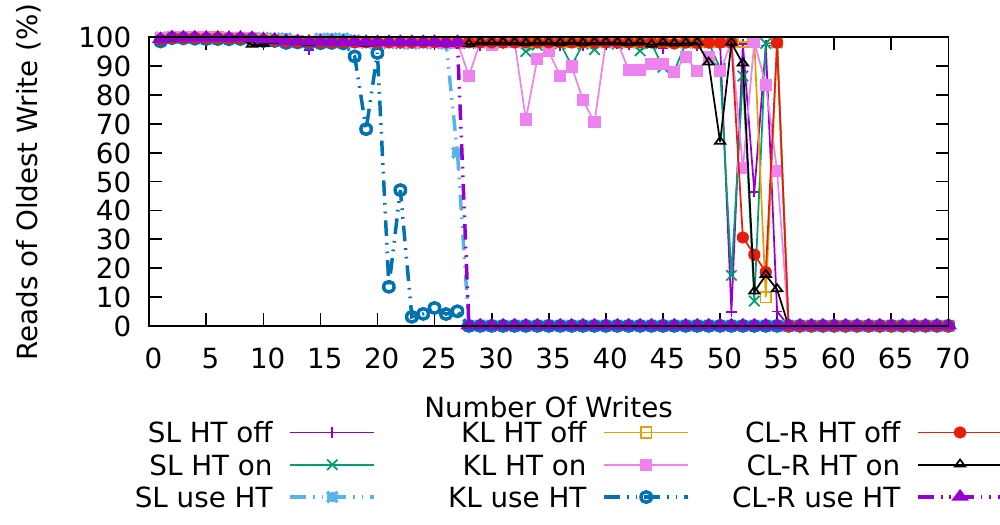}
	\caption{Measuring the size of the store buffer on Kaby Lake and Coffee
	Lake machines. In the experiment, we perform multiple writes to the
	store buffer and subsequently measure the probability of retrieving
	the value of the first (oldest) store. 
	The results agree with 56 entries in the store buffer
	and with a static partitioning between hyperthreads.
	}
	\label{fig:store-buffer-size}
\end{figure}

We now turn our attention to measuring the size of the store buffer.
Intel advertises that 
Skylake processors have 56 entries in the store buffer~\cite{Mandelblat15}.
We could not find any publications specifying the size of
the store buffer in newer processors, but as both Kaby Lake and Coffee Lake R
are not major architectures, we assume  that the size of the store buffers
has not changed.
As a final experiment in this section, we now attempt to use \AttackName 
to confirm this assumption.
To that aim, we perform a sequence of store operations, each to a different
address.
We then use a faulty load aiming to trigger a WTF optimization and
retrieve the value stored in the first (oldest) store instruction.
For each number of stores, we attempt 100 times at each of the 4096 page offsets, 
to a total of 409,600 per number of stores.
\cref{fig:store-buffer-size} shows the likelihood of triggering the WTF
optimization as a function of the number of stores for each of the processor
and configurations we tried.
We see that we can trigger the WTF optimization provided that the sequence
has up to 55 stores. 
This number matches the known data for Skylake and confirms our
assumption that it has not changed in the newer processors.

The figure further shows that merely enabling hyperthreading does not
change the store buffer capacity available to the process. 
However, running code on the second hyperthread of a core 
halves the available capacity, even if the code does not perform any
store.
This confirms that the store buffers are statically partitioned between the
hyperthreads~\cite{Intel_64_IA-32:aorm}, and also shows
that partitioning takes effect only when both hyperthreads are active.

\section{Using \AttackName to Break Kernel Isolation}\label{sec:kernel}
In this section, we show that \AttackName can leak information from
the OS kernel to unprivileged users.  
We first explore a contrived scenario where a dedicated kernel module 
writes data without doing any useful computation.
We then proceed to a more realistic scenario and show leakage from 
real code running inside the kernel.

\subsection{Leaking Memory Writes from the Kernel}
Our proof-of-concept implementation consists of two components.
The first is a kernel module that writes to a predetermined
virtual address in a kernel page.
The second is a user application that 
performs a faulty load from an address in a user page, such that the
page offset of this address the same as the page offset
the kernel module writes to.
Exploiting the WTF
optimization, the user application can retrieve the data written by the
kernel.  We now proceed to describe both parts of our proof-of-concept
implementation. 

\parhead{The Kernel Module.} 
Our kernel module performs a sequence of write operations each to a different page
offset in a different kernel page. 
These pages, like other kernel pages, are not directly accessible to user code.
On older processors, such addresses may be accessible indirectly via Meltdown.
However, we do not exploit this and assume that the user code does not or cannot
exploit Meltdown.

\parhead{The Attacker Application.} 
The attacker application aims to
retrieve kernel information that would normally be inaccessible outside the kernel.
The attacker code first uses \texttt{mprotect} to revoke access to a page.
It then invokes the kernel module to perform the kernel writes.
When the kernel module returns, the attacker performs a faulty load
from the protected page, before transiently leaking the value through
a covert cache channel.

\parhead{Increasing the Window for the Faulty Load.} 
To increase the time window for the faulty load,
our attacker code further delays processing the kernel
store by performing a sequence of store operations before invoking the
kernel module.
Store buffer entries are processed and stored in the cache in
program order~\cite{IslamMBKGES19,abramson1998method,abramson2002method, HilyZH09}.
Thus, filling the store buffer delays processing of later stores.
We further increase the effect of these store operations by first
flushing the addresses they write to from the cache.

 \begin{figure}[t!]
	\centering	
	\includegraphics[page=2,width=\linewidth]{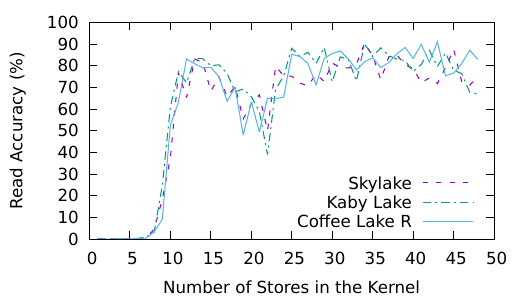}	
	\caption{Probability of recovering kernel values from user space as a function of the number of kernel stores. }
	\label{fig:cross-kernel-eval}
\end{figure}

\parhead{Experimental Evaluation.}
We measure the number of stores that the kernel needs to perform
for \AttackName to be able to recover a value it stores
before returning user space. We use our three Intel machines
with a fully updated Ubuntu 16.04, 
keeping  the kernel mapped in the process's address space.
\cref{fig:cross-kernel-eval} shows the results of our evaluation, where each
experiment is repeated 409,600 times, 100 at each possible page offset.
As the figure shows, after about 10 kernel writes 
the attacker can 
use \AttackName to recover a value written by the kernel on
both machines with about 80\% probability. 

On processors vulnerable to Meltdown, leaving the kernel mapped in the
process's address space disables KPTI, allowing Meltdown attacks on the kernel.
For the Coffee Lake R processor, which includes hardware countermeasures for Meltdown,
KPTI is disabled by default.
In particular, the experiments for this processor in
\cref{fig:cross-kernel-eval} are with the default Ubuntu
configuration.
Ironically, this means that the
hardware countermeasures in Intel's latest CPU generations 
make them more vulnerable to \AttackName.

\subsection{Attack on the AES-NI Key Schedule}
The attack we describe above assumes the most favorable scenario for the attacker.
We now proceed to a more realistic scenario where we show that when
the kernel processes a secret encryption key it may leak enough
information for the user process to recover the key.

The Linux Kernel cryptography API
supports several standard cryptographic schemes that are available to
third-party kernel modules and device drivers which need cryptography. 
For example, the Linux key
management facility and disk encryption services, such as  eCryptfs~\cite{Halcrow05}, 
heavily rely
on this cryptographic library. 

To show leakage from the standard cryptographic API, we implemented
a kernel module that
uses the library to provide user applications with an encryption oracle.
We further implemented a user application that uses the kernel module.
The AES keys that the kernel module uses are only stored in the kernel and
are never shared with the user.
However, our application exploits \AttackName to leak these keys from the kernel.  
We now describe the attack in further details.

\parhead{AES and AES-NI.}
A 128-bit AES encryption or decryption operation consists of 10 \emph{rounds}.
The AES key schedule algorithm expands the AES master key to generate a
separate 128-bit \emph{subkey} for each of these rounds.
An important property of the key scheduling algorithm is that it is reversible.
Thus, given a subkey, we can reverse the key scheduling algorithm to recover the 
master key. For further information on AES, see~\cite{AES}.

Because encryption is a performance-critical operation and to protect against
side-channel attacks~\cite{OsvikST06}, recent Intel processors implement
the AES-NI instruction set~\cite{Gueron10}, which provides instructions that perform
parts of the AES operations.
In particular the \instr{aeskeygenassist} instruction performs part of the
key schedule algorithm.

\begin{lstlisting}[language=C,style=customc, float=htb, label={lst:aeskeysched}, caption={AES-NI Key Schedule},escapechar=|]
aeskeygenassist $0x1, %xmm0, %xmm1|\label{line:keygen1}|
callq  <_key_expansion_128>
aeskeygenassist $0x2, %xmm0, %xmm1|\label{line:keygen2}|
callq  <_key_expansion_128>
...
<_key_expansion_128>:|\label{line:keyexpand}|
pshufd $0xff,%xmm1,%xmm1
shufps $0x10,%xmm0,%xmm4
pxor   %xmm4,%xmm0
shufps $0x8c,%xmm0,%xmm4
pxor   %xmm4,%xmm0
pxor   %xmm1,%xmm0
movaps %xmm0,(%r10)|\label{line:keystore}|
add    $0x10,%r10|\label{line:advance}|
retq   
\end{lstlisting}

\parhead{Key Scheduling in Linux.}
The Linux implementation stores the master key and the 10  subkeys 
in consecutive memory locations.
With each subkey occupying 16 bytes, the total size of the expanded key is 176 bytes.
Where available, the Linux Kernel cryptography API uses AES-NI for implementing
the AES functionality. Part of the code that performs key scheduling for 128-bit
AES appears in \cref{lst:aeskeysched}.
\cref{line:keygen1,line:keygen2} invoke \instr{aeskeygenassist} to perform
a step of generating a subkey for a round.
The code then calls the function \texttt{\_key\_expansion\_128}, which
completes the  generation of the subkey.
The process repeats ten times, once for each round. (To save space we only show two rounds.)

\texttt{\_key\_expansion\_128} starts at \cref{line:keyexpand}.  It performs
the operations needed to complete the generation of a 128-bit AES subkey.
It then writes the subkey to memory (\cref{line:keystore}) before
advancing the pointer to prepare for storing the next subkey (\cref{line:advance})
and returning.

\parhead{Finding the Page Offset.}
Our aim is to capture the key by leaking the values stored in \cref{line:keystore}.
For that, the user application repeatedly invokes the kernel interface
that perform the key expansion as part of setting up an AES context.
Because the AES context is allocated dynamically, its address depends on the
state of the kernel's memory allocator at the time the context is allocated.
This prevents immediate use of \AttackName because the attacker does not know
where the subkeys are stored.

\begin{figure}[t!]
 	\centering
 	\includegraphics[width=.98\linewidth]{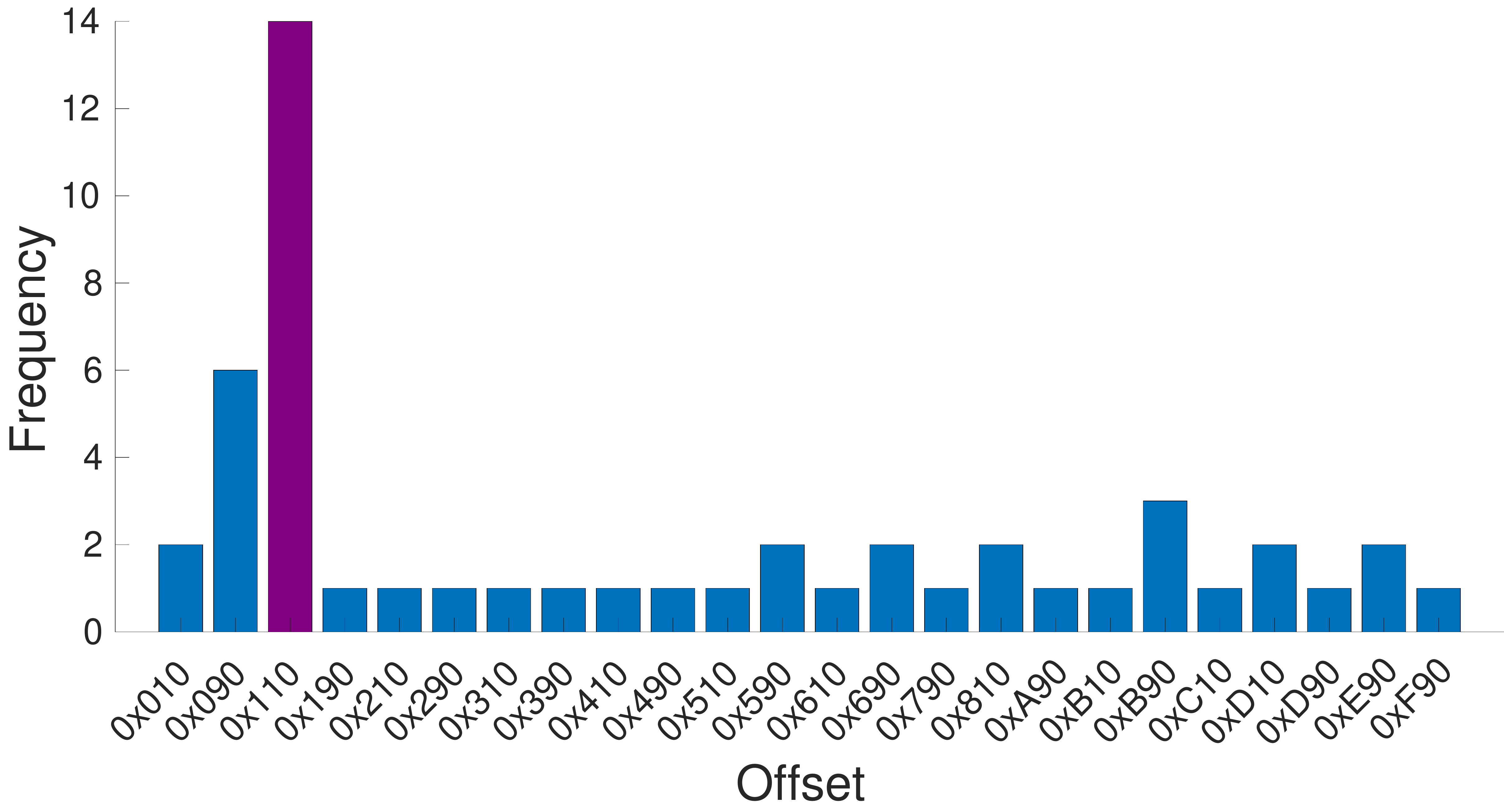}
	\caption{Frequency of observed leaked values. We note that offset 0x110 shows
	more leakage then others.  Confirming against the ground truth, we find that
	all the leaked values at that offset match the subkey byte.}
 \label{fig:rkeyindex}
\end{figure}

We use \AttackName to recover the page offset of the AES context.
Specifically, the user application scans page offsets.
For each offset it
asks the kernel module to
initialize the AES context.  It then performs a faulty load from a protected
page at the scanned offset, and checks if any data leaked.
To reduce the number of scanned offsets, we observe that, as described above,
the size of the expanded key is 176 bytes.  Hence, we can scan at offsets that
are 128 bytes apart and have the confidence that at least one of these offsets
falls within the expanded key.
Indeed, running the attack for 5 minutes we get \cref{fig:rkeyindex}.
The figure shows the number of leaked values at each offset over the full five minutes.
We note the spike at offset 0x110.
We compare the result to the ground truth and find that the expanded key indeed
falls at offset 0x110.  We further find that the leaked byte matches the
value at page offset 0x110.

\parhead{Key Recovery.}
Once we find one offset within the expanded key, we know that neighboring offsets
also fall within the expanded key and we can use \AttackName to recover the
other key bytes.  We experiment with 10 different randomly selected keys and
find that we can recover the 32 bytes of the subkeys of the two final rounds
(rounds~9 and~10) without errors within two minutes.
Reversing the key schedule on the recovered data gives us the master key.

\section{Using \AttackName to Break KASLR}\label{sec:kaslr}
We now show how \AttackName can be used to break Kernel Address
Space Layout Randomization (KASLR). 

\subsection{KASLR Background}
Code injection attacks are a type of vulnerability where the attacker injects
code to the address space of the victim and subsequently
diverts the victim's control flow to execute the injected code. 
A common protection for such attacks is 
to adopt a policy where memory pages are either writable or executable,
but never both.

\parhead{ROP and Return-to-Libc Attacks.}
Return-to-libc~\cite{Solar97} and return oriented programming (ROP)~\cite{Shacham07}
are two related techniques that reuse existing code for exploiting memory
corruption vulnerabilities.
In a nutshell, by overwriting the stack, the attacker can hijack the
control flow, and direct execution into \emph{gadgets} that exist in
the victim's code or in linked libraries.
\cite{Shacham07} demonstrates that a typical library contains enough
gadgets that, when threaded, can perform arbitrary computation.

\begin{figure}[t!]
	\centering
	\includegraphics[width=0.95\linewidth]{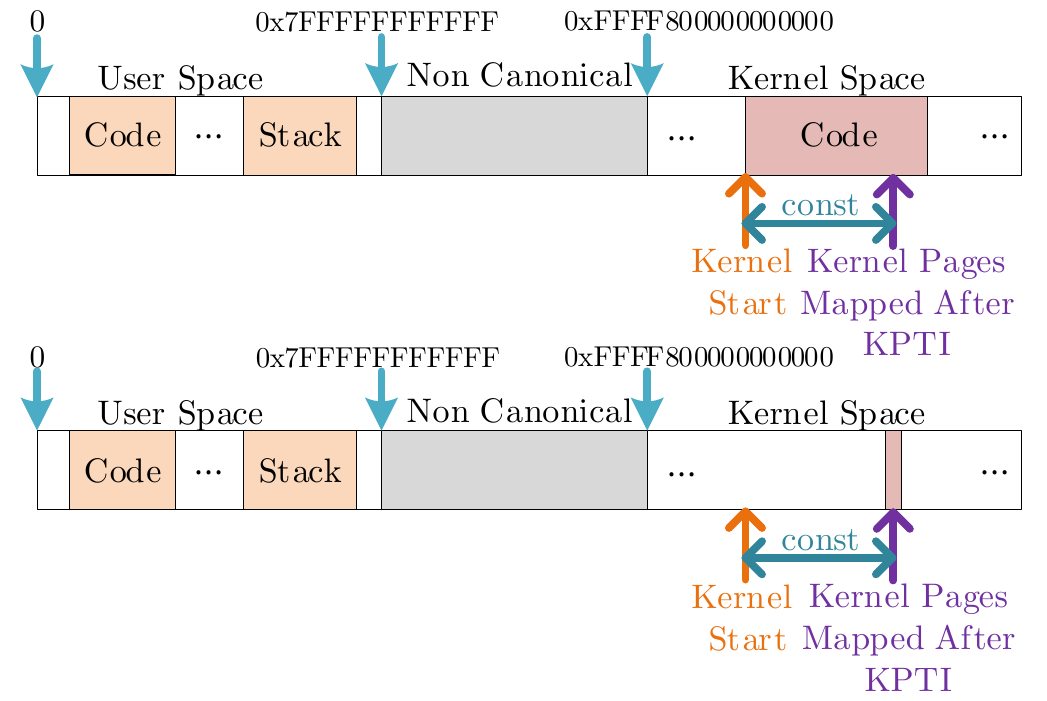}
	\caption{(Top) Address space layout with KASLR but without KPTI.
	(Bottom) User address space with KASLR and KPTI. Most of the kernel is not mapped in the process's address space anymore. 	
	\label{fig:aslr-layout}
	}
\end{figure}

\parhead{ASLR.} 
Address Space Layout Randomization (ASLR) is a probabilistic countermeasure
for ROP.  
The main idea is to introduce
randomness the in the victim memory layout, hiding it from the attacker.
That is, when a process is initialized, ASLR randomizes the locations of the code
and the data (see \cref{fig:aslr-layout}~(top)). 
With ASLR, the attacker needs to find the addresses of code gadgets
to be able to use them. 

\parhead{KASLR on Linux Systems.} 
On Linux systems, KASLR had been supported since kernel version 3.14 and
enabled by default since around 2015. As 
\cite{jang2016breaking} note, the amount of entropy present depends on the kernel
address range as well as on the alignment size which is usually  multiple of
page size. 

\parhead{KASLR and KPTI.}
As a countermeasure to the Meltdown attack~\cite{Lipp2018meltdown}, OSs running
on Intel processors up to the latest Coffee Lake architecture have deployed the
Kernel Page Table Isolation (KPTI) mechanism, which removes the kernel
from the address space of user processes (see \cref{fig:aslr-layout}~(bottom)).
To allow the process to switch to the kernel address space, the system
leaves at least one kernel page in the address space of the user process.
Because the pages required for the switch do not contain any secret information, there is no need
to hide it from Meltdown.

The KPTI patch is based on KAISER~\cite{Gruss2017KASLR}, which was
originally designed to protect the kernel from side-channel
attacks that break KASLR~\cite{Gruss2016Prefetch,Hund2013,jang2016breaking}.
We now proceed to show that \AttackName can reveal the location of the 
kernel entry page left in the user address space,
thereby breaking KASLR.

\subsection{Using \AttackName to Break Kernel ASLR}\label{sec:fallout-aslr}
\parhead{Attack Overview.} 
Our attack is based on the disparity between the effects of causes of faults
(see \cref{table:sbr-table}).
Specifically, we note that when accessing an unmapped kernel page, the WTF
optimization is not triggered and the \AttackName attack fails.
Thus, to perform the attack, we replace the read from  \texttt{attacker\_page} in
\cref{line:spec-trans} with a read from a page within the kernel address range.
When the page we access is mapped, \AttackName succeeds and we retrieve
a value from the store buffer.
Otherwise no value is retrieved from the store buffer.

\parhead{Experimental Setup.}
We evaluate \AttackName on two Intel machines, a Kaby Lake i7-7600U and a
Coffee Lake R i9-9900K. Both machines run a fully updated Ubuntu 16.04
system, with all countermeasures in their default configuration. On both
systems, we empirically test the possible locations on the kernel in its
address space obtaining about 490 locations, implying about 9 bits of entropy.

\parhead{Experimental Results.}
We run the attack 1000 times each, on both the Kaby Lake and the Coffee
Lake machines. Our attack can recover the kernel location
with 100\% accuracy on both machines, within about 0.27 seconds.

\section{Transient Execution and Microcode Assists}\label{sec:no-supress}
Recall (\cref{sec:speculative-attacks}) that \cite{canella2018systematic} 
classifies transient execution attacks based on the cause of transient execution.
Spectre-type attacks are caused by misprediction of data or control flow,
whereas Meltdown-type attack are caused by transient execution beyond a fault.
We now investigate \emph{microcode assists}, a microarchitectural mechanism
that has not yet been explored in the context of transient execution
attacks.  We identify \muop redispatching, which occurs as part of invoking
microcode assists, as a new cause for transient execution that extends the
classification of \cite{canella2018systematic}.

\subsection{Microcode Assists}
\muops are typically implemented in hardware.  
However, when complex processing is required for rare corner
cases, it may not be cost effective to implement some of the functionality
in hardware.
Instead, if such a case occurs during the execution of a \muop,
the \muop is \emph{redispatched}, \ie sent back to the dispatch queue
for execution, together with a \emph{microcode assist}, a microcode procedure 
that handles the more complex scenario.

The Intel optimization manual~\cite{Intel_64_IA-32:aorm} lists two scenarios in
which microcode assists are invoked: when handling subnormal floating point numbers
and in some cases during the processing of the \instr{vmaskmov} (Conditional SIMD Packed 
Loads and Stores) instruction. \cite{CostanD16} lists further scenarios.

In this work we are interested in microcode assists that occur as part of the
virtual to physical address translation.
While Intel does not publish official documentation on the process, 
it has applied for a related patent~\cite{glew1997method},
on which we base our discussion.
We only describe the parts of the patent that are relevant to this work.
We refer the reader to the patent application for a more complete description.

When the processor handles a memory access (load or store) it
needs to translate the virtual address specified by the program to
the corresponding physical address.
For that, the processor first consults the Data Translation Look-aside Buffer (DTLB),
which caches the results of recent translations.
In the case of a page miss, \ie when the virtual address is not found in the
DTLB, the \emph{page miss handler} (PMH) attempts to consult the page map to find
the translation.
In most cases this translation can be done while the \muop is speculative.
However, in some cases the page walk has side effects that cannot take place
until the \muop retires.
Specifically, store operations should mark pages as dirty and all memory
operations should mark pages as accessed.
Performing these side effects while the \muop is speculative risks generating
an architecturally-visible side effect for a transient \muop. (Recall that the processor cannot
determine whether speculative \muops will retire or not.)
At the same time, recording all the information required for setting the
bits on retirement would require a large amount of hardware that will only
be used in relatively rare cases.
Thus, to handle these cases, the processor redispatches the \muop and arranges
for a microcode assist to set the bits when the \muop
retires.

\subsection{\AttackName and Microcode Assists}
To test the effects of microcode assists on \AttackName, 
Use the code in \cref{lst:pseudocode-assist}.
The code is basically the same as \cref{lst:pseudocode},
except that we use SGX-Step~\cite{BulckPS17}
to replace the call to
\texttt{mprotect} and instead mark \texttt{attack\_page} as not accessed (\cref{line:clear_access}).
Furthermore, because microcode assists do not generate faults, we do not need fault suppression,
and remove the TSX transaction.

\begin{lstlisting}[language=C,style=customc, float=htb, label={lst:pseudocode-assist}, caption={Pseudocode of \AttackName with microcode assists.  Note that no fault suppressison is required.},escapechar=|]
char* victim_page = mmap(..., PAGE_SIZE, ...);|\label{line:mmap-target-a}|
char* attacker_page = mmap(..., PAGE_SIZE, ...);|\label{line:mmap-attacker-a}|

offset = 7; |\label{line:target-address-a}|
victim_page[offset] = 42; |\label{line:target-write-a}|

clear_access_bit(attacker_page);|\label{line:clear_access}|
memory_access(lut + 4096 * attacker_page[offset]); |\label{line:spec-trans-a}|

for (i = 0; i < 256; i++) {|\label{line:probe-start-a}|
  if (flush_reload(lut + i * 4096)) { |\label{line:reload-a}|
    report(i);
  }
}|\label{line:probe-end-a}|
\end{lstlisting}

\parhead{Recovering the Leaked Data.}
As in \cref{sec:toy-read} we use \FlushReload to recover the leaked data.
We repeat the experiment on three processor generations: Skylake, Kaby Lake, 
and Coffee Lake R.
In all architectures reading from the entry in the probe array corresponding
to the value 42 has a short access time.

\subsection{Assist-based vs.\ Meltdown-type}
\cite{canella2018systematic} list several properties of Meltdown-type attacks.
Assist-based transient execution shares \emph{some} properties with
Meltdown.  Specifically, it relies on deferred termination of a \muop to
bypass hardware security barriers and attacks based on it can be mitigated 
by preventing the original leak.
However, unlike Meltdown-type techniques, assists do not rely on faults.
Consequently, no fault suppression techniques are required.

\section{Countermeasures}

\parhead{Flushing-Based Countermeasures.} 
Because the store buffer is not shared across hyperthreads, 
leaks can only occur when the security domain changes within
a hyperthread.
Thus, flushing the store buffer on security domain change is
sufficient to mitigate the attack.
In particular, we verified that using \instr{mfence} as part of the switch from kernel mode to
user mode thwarts the attack.

\parhead{Limitations.}
As mentioned above, the attacks described in \cref{sec:kernel} are unable to
leak information across hyperthreads . Moreover, as Meltdown software countermeasures
(KPTI) flush the buffer on leaving the kernel, and as the store buffer is
automatically flushed on change of the \texttt{CR3} register (i.e., on context
switch), only latest generation Coffee Lake R machines are vulnerable to the
attack described in \cref{sec:kernel}. Ironically, the hardware mitigations
present in newer generation Coffee Lake R machines make them more vulnerable to
\AttackName than older generation hardware.

\section{Conclusion}
With \AttackName, we demonstrate a novel Meltdow-type effect exploiting a previously unexplored microarchitectural component, namely the store buffer.
The attack enables an unprivileged attacker to leak recently written values from the operating system.
Furthermore, we demonstrate how \AttackName allows to break kernel ASLR with 100\% accuracy within 0.27 seconds.
While \AttackName affects various processor generations, we showed that also recently introduced hardware mitigations are not sufficient and futher mitigations need to be deployed.

\section*{Acknowledgments}
This research was supported in part by Intel Corporation.
The research presented in this paper was partially supported by the Research Fund KU Leuven.
Jo Van Bulck is supported by a grant of the Research Foundation -- Flanders (FWO).
The project was supported by the European Research Council (ERC) under the European Union's Horizon 2020 research and innovation programme (grant agreement No 681402).
It was also supported by the Austrian Research Promotion Agency (FFG) via the K-project DeSSnet, which is funded in the context of COMET – Competence Centers for Excellent Technologies by BMVIT, BMWFW, Styria and Carinthia.
Additional funding was provided by a generous gift from Intel.
Researchers from Worcester Polytechnic Institute are supported by National Science Foundation under the grant CNS-1618837 and CNS-1814406. 

Any opinions, findings, and conclusions or recommendations expressed in this paper are those of the authors and do not necessarily reflect the views of the funding parties.

{\footnotesize \bibliographystyle{plain}
\bibliography{bibliography}}

\end{document}